\documentclass[%
 reprint,
 amsmath,amssymb,
 aps,
longbibliography,
noeprint
]{revtex4-1}

\usepackage{graphicx}
\usepackage{dcolumn}
\usepackage{bm}
\usepackage{bbold}
\usepackage{color}

\renewcommand{\vec}{\mathbf}
\newcommand{\Id}{\mathbb{1}}

\usepackage[colorlinks=true,urlcolor=blue,linkcolor=blue,citecolor=blue]{hyperref}

\begin{document}


\title{Chiral topological phases in designed mechanical networks}

\author{Henrik Ronellenfitsch}
\email{henrikr@mit.edu}
\author{J\"orn Dunkel}
\email{dunkel@mit.edu}
\affiliation{Department of Mathematics, Massachusetts Institute of Technology, 77 Massachusetts Avenue, Cambridge, MA 02139, U.S.A.}

\date{\today}

\begin{abstract}
Mass-spring networks (MSNs) have long been used as approximate descriptions of 
many biological and engineered systems, from actomyosin networks
to mechanical trusses. In the last decade, MSNs have re-attracted theoretical interest as models for phononic metamaterials with exotic properties such as negative Poisson's ratio, negative effective mass, or gapped vibrational spectra.
A practical advantage of MSNs is their tuneability, which allows the inverse design of  materials with pre-specified bandgaps. Building on this fact, we demonstrate here that designed MSNs, when subjected to Coriolis forces, can host topologically protected chiral edge modes at predetermined frequencies, thus enabling robust unidirectional transmission of mechanical waves. Similar to other recently discovered topological materials, the topological phases of MSNs can be classified by a Chern invariant related to time-reversal symmetry breaking.
\end{abstract}

\maketitle
\section{Introduction}
Topological mechanics~\cite{Susstrunk2016} is a rapidly growing research field that studies classical analogs of topological effects in quantum many-body physics~\cite{Asboth2016}. A prime example are spectrally gapped mechanical systems that can host topologically protected zero modes at their boundaries~\cite{Kane2014,Sussman2016,Susstrunk2015}, similar to localized electronic excitations in the quantum spin Hall effect~\cite{Kane2005}.
Another important class of examples are solid- or fluid-mechanical systems with broken time-reversal symmetry, which can  exhibit chiral edge modes at finite frequency ~\cite{Souslov2016,Nash2015,Mitchell2018,Wang2015b}, analogous to the
(anomalous) quantum Hall effect~\cite{Nagaosa2010,Hasan2010}.
Because these edge modes are topologically protected and robust against
the introduction of defects, they may provide a powerful tool for the 
resilient localized transmission of sound signals in elastic materials~\cite{Xia2017}.

\par
Over the last five years, substantial progress has been made in the understanding of topological phenomena in a wide variety of classical systems, ranging from mechanical systems with lattice symmetry inspired by quantum analogs~\cite{Wang2015c} and amorphous networks~\cite{Agarwala2017,Mitchell2018} to  active systems~\cite{Souslov2016,Shankar2017,Woodhouse2018}, electrical circuits~\cite{Lee2017,Imhof2017,Kotwal2019}, and even ocean 
waves~\cite{Delplace2017}. Many of the recently discovered mechanical topological insulators rely on a known underlying lattice structure~\cite{Wang2015c,Souslov2016} or curvature~\cite{Delplace2017} to induce the required gaps in their excitation spectra. From a practical perspective, it would be interesting to design and build more general structures with desired topological properties. 
\par
Complementing recent work aimed at engineering continuum topological
insulators~\cite{Christiansen2019}, we consider here the design of topological excitations in bandgap-optimized~\cite{Ronellenfitsch2018} mass-spring networks (MSNs).
Specifically, we will demonstrate that MSNs with an inversely designed bandgap can
host topologically protected finite-frequency edge modes, and
convert non-robust non-topological edge modes into robust topological edge modes when time-reversal symmetry is broken.
While many traditional topological materials, including those based on a hexagonal
lattice like the Haldane model~\cite{Haldane1988}, do not possess the mode conversion property, this desirable feature is frequently encountered in our designed MSNs.

\par
In the remainder, we focus on the dynamics of periodic crystals of 2D mechanical balls-and-springs networks. In all cases, the spring stiffnesses of these MSNs were numerically tuned such that the excitation spectrum exhibits a band gap (using the algorithm introduced in Ref.~\cite{Ronellenfitsch2018}). In formal analogy with quantum Hall systems~\cite{Nagaosa2010,Hasan2010}, we will then break time reversal symmetry by placing the MSN into a rotating frame, with the Coriolis forces playing the role of an external magnetic field. To study and characterize the topological phase transition and the emerging protected chiral edge modes in detail, we will \emph{(i)}~numerically calculate the non-zero Chern invariant associated
to the topological phase, \emph{(ii)}~demonstrate the dynamics
of the localized chiral edge excitations in numerical simulations, and \emph{(iii)}~explicitly identify those dynamical edge modes that are related to topological protection.

\section{Dynamics of mechanical networks}

MSNs provide a generic modeling framework for many physical systems. The potential energy of
an MSN with $E$ springs is given by
\begin{align*}
    V = \frac{1}{2} \sum_{e=1}^E k_e \left( \ell_e - \ell^{(0)}_e \right)^2,
\end{align*}
where $k_e$ is the stiffness of spring $e$, $\ell_e$ is its current length
and $\ell^{(0)}_e$ is its preferred rest length. Here, we are interested in the dynamics near the equilibrium configuration
where all springs are at their rest lengths, $\ell_e=\ell^{(0)}_e$, corresponding to the masses being at positions $\mathbf{x}_i^{(0)}$. Expanding in small deviations $\mathbf{u}_i = \mathbf{x}_i
-\mathbf{x}_i^{(0)}$, we obtain the linearized equations of motion,
\begin{align*}
    m \ddot{\mathbf{u}} + K\mathbf{u} = 0,
\end{align*}
where $K$ is the stiffness matrix of the network and $m$ is the mass
of the balls (we assume identical masses throughout).
The vector $\mathbf{u}$ generally has $dN$ components, where $d$ is the dimension of
space and $N$ is the number of masses. From now on, we specialize to the case $d=2$.
The stiffness matrix can be further decomposed as $K= Q\hat{k}Q^\top$, where
$Q$ is the equilibrium matrix encoding the network geometry and
$\hat{k}=\operatorname{diag}(k_1, k_2,\dots, k_E)$ is the diagonal matrix
of spring stiffnesses~\cite{Lubensky2015}.

\par
The MSN dynamics, specifically its harmonic response and
its phononic modes, are encoded in the eigenmodes

\begin{align}
    K\mathbf{u}_j = m\omega_j^2 \mathbf{u}_j,
    \label{eq:eigen}
\end{align}
where $\omega_j$ are the eigenfrequencies. If the network is a crystal consisting of $N_c$ periodically repeated unit cells
with lattice vectors $\vec R_{\ell}$,
the dynamical problem can be simplified by performing a lattice Fourier
transform~\cite{Lubensky2015},
\begin{align*}
  u_{n}(\vec R_\ell) &= \frac{1}{N_c} \sum_{\vec k} e^{-i\vec k \cdot \vec x_{i}} u_n(\vec k) \\
  u_{n}(\vec k) &= \sum_{\vec R_\ell} e^{i\vec k \cdot \vec x_{i}} u_n(\vec R_\ell),
\end{align*}
where we decompose the rest positions $\mathbf{x}_i = \mathbf{R}_\ell + 
\mathbf{v}_n$. Here, $\ell$ indexes the unit cell and $n$ indexes
the degree of freedom within the unit cell. The wavevector \mbox{$\mathbf{k} = \sum_{i=1}^d \frac{b_i}{N_i}
\mathbf{K}_i$} with $b_i\in \mathbb{Z}$ lies in the first Brillouin zone, $N_i$
is the number of unit cells in the $i$th dimension, and
the reciprocal lattice vectors satisfy $\mathbf{K}_i \cdot \mathbf{R}_j = 2\pi
\delta_{ij}$ for the primitive lattice vectors $\mathbf{R}_j$.
The Fourier transform decouples the eigen-problem Eq.~\eqref{eq:eigen}
for different wavevectors $\mathbf{k}$ and leads to a phononic band structure~$\omega_n(\mathbf{k})$.
\par
The band structure~$\omega_n(\mathbf{k})$ is of great interest both scientifically and
for engineering applications because it efficiently encodes the elastic
response of the infinite network. Specifically, band structure engineering
allows for explicit tuning of wave propagation in acoustic materials
and can be used to design, for instance, waveguides, acoustic cloaks, or
selective sound suppression.
Whereas in a generic band structure, the shape and
frequency of the acoustic modes depends strongly on the details
of the dynamics, topological modes are protected by
an integer invariant, which cannot change through continuous
changes of the interaction parameters. 
\par
Although the physical realizations of topological insulators are vast already in the quantum case~\cite{Asboth2016,Hasan2010},
the possible invariants and topological phases have been  completely classified~\cite{Altland1997}.
In linear topological mechanics, a similar scheme exists as long as the dynamical matrix is positive definite~\cite{Susstrunk2016}.
In the following, we shall focus on one particular class
of classical topological band structures for two-dimensional systems.

\section{Planar Chern insulators}

\begin{figure*}
    \centering
    \includegraphics[width=.98\textwidth]{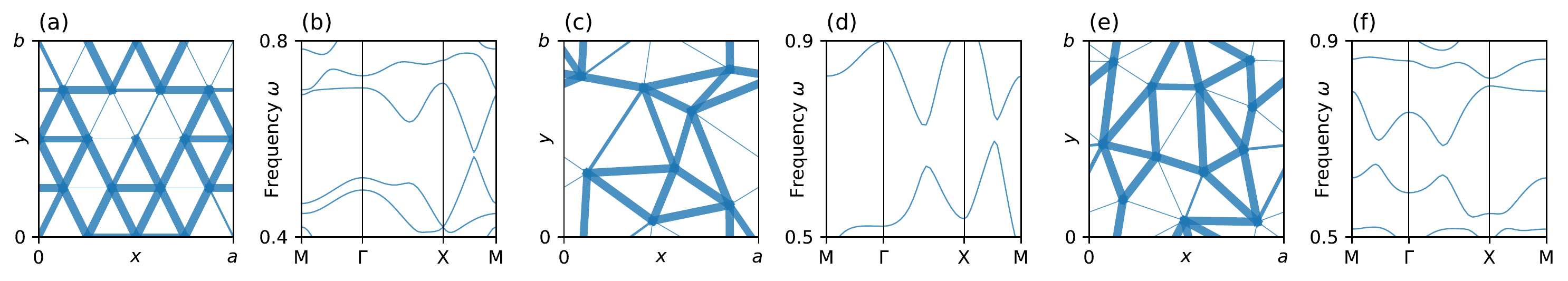}
    \caption{\textbf{Bandgap tuned networks and topological band structures.}
    (a)~Tuned spring stiffnesses (indicated by line thickness) on the basis of triangular
    grids with $4\times 4$ unit cells were tuned to have band gaps.
    (b)~The final band structures exhibit band shapes reminiscent of band inversion.
    (c,e)~Tuned network on top of a topology based on
    the Delaunay triangulation of a randomized point set. (d,f)~The band
    structure again appears to exhibit band inversion.
    All three networks undergo a topological phase transition as 
    time-reversal symmetry is broken.}
    \label{fig:1}
\end{figure*}
Topological band structure is intimately related to the theory of Berry phases, or geometrical phases~\cite{Berry1984}.
While a full account of the underlying theory is beyond the scope of this paper,
the fundamental result can be stated for a linear dynamical system
$i\dot {\boldsymbol{\psi}} = H(\mathbf{r})\boldsymbol{\psi}$ with Hermitian matrix  $H(\mathbf{r})$
which depends on some parameter $\mathbf{r}$.
If the system is prepared in an instantaneous eigenstate 
$H(\mathbf{r})\mathbf{u}(\mathbf{r}) =
\lambda(\mathbf{r}) \mathbf{u}(\mathbf{r})$ and the parameter $\mathbf{r}$ is varied
adiabatically along a closed curve $\mathcal{C}$ in parameter space,
then the solution will always remain in the instantaneous eigenstate. After traversing the curve, the solution will pick up a phase factor $e^{i\gamma_{\mathcal{C}}}$
with
\begin{align}
    \gamma_{\mathcal{C}} = \int_{\mathcal C} d\mathbf{r} \cdot \mathbf{A}(\mathbf{r}),
    \quad \mathbf{A}(\mathbf{r}) = i \mathbf{u}(\mathbf{r})^H 
        \nabla_{\mathbf{r}} \mathbf{u}(\mathbf{r}).
        \label{eq:berry}
\end{align}
This is the celebrated Berry phase with $\mathbf{A}(\mathbf{r})$ the Berry connection (superscript $H$ denotes the  Hermitian transpose).
While the Berry connection changes under reparametrizations of the curve 
(gauge transformations), the phase is invariant up to $2\pi$, and therefore
in principle a physical observable.
One particular parameter space of interest is the Brillouin zone of a crystal.
In two dimensions, the BZ has the topology of a torus, such that any curve connecting
$\mathbf{k}$ and $\mathbf{k} + \mathbf{K}$ is closed (because wavevectors
$\mathbf{k}$ and $\mathbf{k}+\mathbf{K}$ are equivalent if $\mathbf{K}$
is a reciprocal lattice vector). By Stokes' theorem,
Eq.~\eqref{eq:berry} can then be expressed as a surface integral independent
of the curve,
\begin{align}
    \gamma_{\mathcal{C}}  = \int_{\mathrm{BZ}} d\mathbf{k} \, \Omega(\mathbf{k})\equiv \chi,
    \label{eq:chern}
\end{align}
where $$\Omega(\mathbf{k}) = \partial_{k_1} A_2(\mathbf{k}) - \partial_{k_2} A_1(\mathbf{k})$$
is called the Berry curvature.
Equation~\eqref{eq:chern} defines the \emph{Chern number}~$\chi$,
which is an integer modulo $2\pi$, and characterizes the eigenstates $\{\mathbf{u}(\mathbf{k})\}_{\mathbf{k}\in \mathrm{BZ}}$.
Thus, because in a crystal each eigenstate parametrized by
the wavevector $\mathbf{k}$ corresponds to a
band, it is possible to assign a topological Chern number~$\chi_n$ to each band~$n$. This Chern number does not change under perturbations of the
matrix $H(\mathbf{k})$, unless bands cross. Then, the eigenstates are no longer non-degenerate and the above analysis fails.

The Chern number defined by Eq.~\eqref{eq:chern} is nonzero only if the dynamics are not time-reversal invariant. If the system has time reversal invariance, $\Omega(\mathbf{k})$ is an odd function of $\mathbf{k}$, and the integral over the
Brillouin zone vanishes.

For systems with many bands and a gap between
bands $n'$ and $n'+1$, the key insight~\cite{Hatsugai1993} is then that 
one can associate an invariant to the \emph{gap} itself, namely
\begin{align*}
    C(n') = \sum_{n \leq n'} \chi_n,
\end{align*}
which can only change if the gap closes due to a perturbation of~$H(\mathbf{k})$. The gap-Chern number $C$ characterizes the bulk of a gapped crystal.
Near a boundary to another gapped crystal with a different $C$ 
or to the vacuum,
the topology of the system must therefore change locally by closing 
the gap. This argument implies the existence of modes that are localized
to the boundary between different topological phases and located in
the bulk gap.
Because these modes are tied to the bulk topological invariants,
they are robust and must always exist, regardless of the specific shape
of the boundary. We note that while for historical reasons the notion of adiabatic changes of parameters was invoked to
define the Chern number, no actual adiabatic
processes are necessary for it to exist, and
it makes sense for any Fourier-transformed
Hamiltonian.

For numerical purposes, the above integrals can be discretized
while retaining their gauge-invariant characteristics~\cite{Fukui2005}.
This way, Chern numbers can be computed robustly and
quickly with reasonably coarse discretizations of
the Brillouin zone. In addition, any Chern number
numerically computed in this way will automatically be an
integer.

In the remainder of this paper, we demonstrate that such topologically
protected edge modes can indeed exist in mechanical networks which have been
tuned to exhibit bandgaps at specified frequencies, opening up an inverse-design pathway towards explicitly programmable topology.

\section{Inverse bandgap design}

There are many mechanical systems that possess
topological gaps by virtue of their lattice structure. 
Here, we consider a different approach by tuning a desired gap into the spectrum of a mechanical network through numerical Linear Response Optimization (LRO)~\cite{Ronellenfitsch2018}. Starting from a basic lattice topology such as a triangular
grid or a randomized unit cell topology defining mass points and springs
(Fig.~\ref{fig:1}), the spring stiffnesses
$k_e$ are numerically optimized to produce a gapped material
between two desired bands. Using the LRO approach introduced in  Ref.~\cite{Ronellenfitsch2018}, we minimize
the average response of the network at frequency $\omega$,
\begin{align}
    R_\omega(\hat{k}) = \operatorname{Tr} \left( G_\omega(\hat{k})^H G_\omega(\hat{k}) \right),
    \label{eq:response}
\end{align}
where $G_\omega(\hat{k}) = \left(m\omega^2\mathbb{1} - Q\hat{k}Q^\top\right)^{-1}$ is the
linear response matrix to harmonic forcing with frequency $\omega$
and $\operatorname{Tr}(\cdot)$ is the matrix trace.
Numerically minimizing Eq.~\eqref{eq:response} over $\hat{k}$
while fixing a certain \mbox{$\omega_n < \omega < \omega_{n+1}$} for eigenmodes $\omega_n$ is then equivalent to maximizing a spectral gap between
the $n$th and $(n+1)$th eigenvalue. 
Generalizing from spectral gaps to bandgaps, since the Fourier
transform is a linear map that block-diagonalizes $G_\omega(\hat{k})$,
the trace in Eq.~\eqref{eq:response} is replaced by a sum over
the traces over the responses at each individual wavevector
$\mathbf{k}$,
$G_\omega(\hat{k}, \mathbf{k})$.
For practical purposes, this sum is truncated, and
only traces over a small number of wavevectors are actually used
in the numerical optimization. To avoid the spring stiffnesses
converging to either zero or infinity, we additionally
impose bound constraints \mbox{$0.1 \leq k_e \leq 1.0$}, and employ
the Limited-memory Broyden--Fletcher--Goldfarb--Shanno algorithm~\cite{Byrd1995}
to perform the numerical optimization.
Particle masses are set to unity ($m=1$).

\par
The above LRO approach generalizes to arbitrary network topologies and dimensions~\cite{Ronellenfitsch2018}. Throughout this paper, we will illustrate general ideas by focusing on three specific examples of bandgap-tuned networks: One with a regular triangular grid
unit cell topology, and two different randomized unit cell
topologies [Fig.~\ref{fig:1}(a,c,e)]. All three networks were optimized  to exhibit a bandgap at some predetermined frequency. Despite some notable differences between them, their band structures all show features reminiscent of band inversion [Fig.~\ref{fig:1}(b,d,f)], a characteristic that is often (but not always) present in topological band structures~\cite{Xi2017,Zhu2012,Fu2007}.
\par
Adopting band inversion as an indicator for the potential
existence of a topological transition, all that remains
to do is to break time-reversal invariance of the system dynamics by introducing a suitable interaction.
In the case of electronic systems, an externally applied magnetic field can provide such a symmetry-breaking  interaction~\cite{Hasan2010}. A classical  counterpart considered in the remainder is the Coriolis force~\cite{Oza2014}  which breaks the time reversal symmetry of the MSN dynamics when the mechanical network is placed in a rotating frame of reference~\cite{Wang2015c}.

\section{Mechanical networks in rotating frames}
\begin{figure*}
    \centering
    \includegraphics[width=\textwidth]{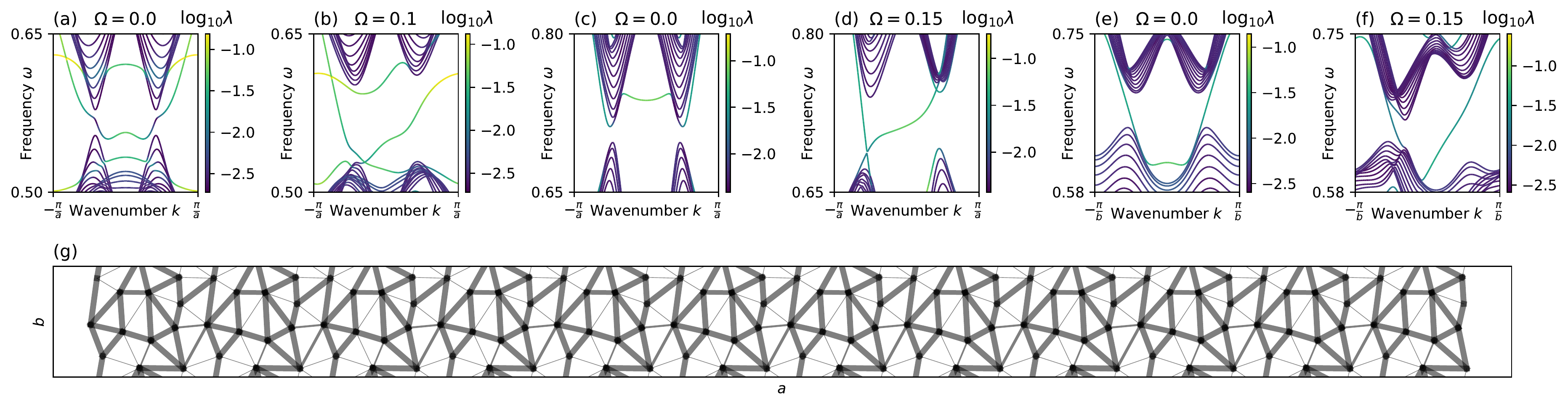}
    \caption{\textbf{Topologically protected edge bands.}
    Considering the same networks as in Fig.~\ref{fig:1}, we constructed ribbon-like 1D crystal realizations that are infinitely periodic in one lattice direction, and of finite extent (12 unit cells wide) with open boundary conditions in the other direction; see example in panel (g), which corresponds to  12 horizontally concatenated units of the network in Fig.~\ref{fig:1}(e) periodically continued along the vertical direction. 
    (a--f)~The resulting one-dimensional band structures consist of bulk bands that are delocalized (dark blue) and localized edge bands (green and yellow). All bands are colored according to the how localized the corresponding modes are, with the localization of a mode $\mathbf{u}(k)$ being measured by the participation ratio $\lambda = (\sum_n |u_n|^4)/(\sum_n |u_n|^2)^2$ where index $n$ runs over all vertices in the unit cell of the ribbon. 
    Panels (a,b) correspond to the optimized network topology in  Fig.~\ref{fig:1}(a); panels (c,d) correspond to the network topology in  Fig.~\ref{fig:1}(c); panels (e,f) correspond to the network topology in Fig.~\ref{fig:1}(e). Generically, networks designed via LRO can and will have localized edge modes even in the topologically trivial regime~$|\Omega|<\Omega_c$, see panels~(a,c,e). However, when the topological phase transition is crossed at some nonzero rotation rate $|\Omega|=\Omega_c$, topologically protected localized bands appear as evident from panels (b,d,f). 
    }
    \label{fig:2}
\end{figure*}
\begin{figure*}[t]
    \centering
    \includegraphics[width=\textwidth]{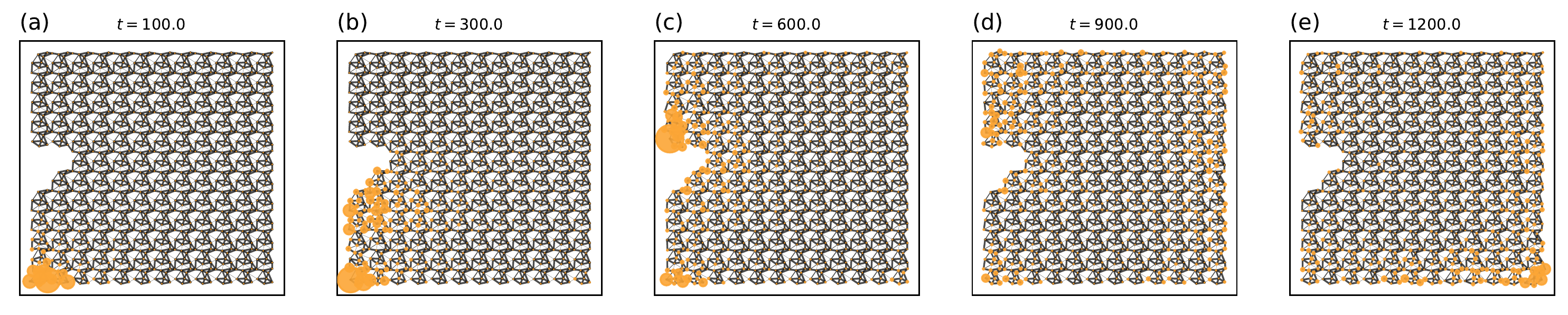}
    \caption{\textbf{Traveling excitation in a rotating mechanical Chern network.} We simulated the dynamics of an MSN consisting of $12\times 12$ unit cells as shown in Fig.~\ref{fig:1}(c)
    and analyzed in Fig.~\ref{fig:2}(c,d). To create an edge defect, a number of unit cells was removed from the left boundary. Simulations of Eqs.~\eqref{eq:dynamics-rot-final-forced} were performed in the topological regime with $\Omega = 0.15$; see also Supplemental Movie 1.
    (a) Between $t=0$ and $t=150$, a single node in the bottom left corner is harmonically forced with a frequency $\omega=0.71$ in the bulk gap; see Fig.~\ref{fig:2}(d).
    (b--e) Starting at $t=150$, a localized excitation travels along the edge of the mechanical Chern insulator, even
    moving over a local perturbation of the boundary.
    }
    \label{fig:3}
\end{figure*}
To sketch the general procedure for formulating the MSN dynamics in a rotating frame, we first consider a point mass in a harmonic potential with stiffness $K$ confined to the $x$--$y$ plane, and under the influence of 
a constant rotation perpendicular to the plane,
$\vec\Omega = (0,0,\Omega)$.
Let $\vec x$ be the position of the point mass as measured from
the rotational axis. Then Newton's equations of motion in the rotating frame take the form
\begin{align}
\label{e:harm_newt}
  \ddot{\vec x} = - K (\vec x - \vec x_0 ) - 2 \vec\Omega \wedge \dot{\vec x} - \vec\Omega \wedge
  (\vec \Omega \wedge \vec x).
\end{align}
The Coriolis force is 
$$- 2 \vec\Omega \wedge \dot{\vec x}
=-2 (0,0,\Omega) \wedge (\dot x, \dot y, 0) = 2\Omega (\Gamma \dot{\vec x'}, 0),$$
where
\begin{align*}
  \Gamma = \begin{pmatrix}
  0 & 1 \\ -1 & 0
\end{pmatrix}
\end{align*}
encodes the cross product and we introduced the 2D vector $\vec x'$.
Similarly, the centrifugal force is
$- \vec\Omega \wedge
(\vec \Omega \wedge \vec x') = \vec\Omega \wedge (\Omega \Gamma \vec x', 0) =
-\Omega^2 (\Gamma^2 \vec x', 0) = \Omega^2 (\vec x', 0)$.
Clearly, the fictitious forces lie in the plane of rotation,
so that from now on we can analyze the system in 2D. Dropping the primes, Eq.~\eqref{e:harm_newt} then yields the in-plane equations of motion
\begin{align}
  \ddot{\vec x} = - K (\vec x - \vec x_0 ) + 2\Omega \Gamma \dot{\vec x}
  + \Omega^2 \vec x. \label{eq:single-particle}
\end{align}

We can now generalize from a single particle to the full MSN dynamics by collecting all the $x$ coordinates of the point masses in the network into the first $N$ components of the $2N$-component vector $\mathbf{x}$, and all the $y$ coordinates into the second $N$ components. Then the matrix $\Gamma$ takes the form
\begin{align*}
  \Gamma = \begin{pmatrix}
  0 & \Id \\
  -\Id & 0
\end{pmatrix},
\end{align*}
where $\mathbb{1}$ is the $N\times N$ identity matrix,
and $K$ now denotes the stiffness matrix such that
Eq.~\eqref{eq:single-particle} remains formally unchanged.

We would like to express these equations in terms of small displacements around
an equilibrium configuration. In doing so, we need to take into account that the equilibrium configuration is changed by the rotation. To find the new equilibrium positions~$\mathbf{x}^*$ in the rotating frame, we set $\ddot{\vec x} = \dot{\vec x} = 0$ and solve for $\mathbf{x}^*$,
\begin{align*}
  (K - \Omega^2 \mathbb{1}) \vec x^* = -K\vec x_0.
\end{align*}
Thus, a steady state exists unless the rotation frequency $\Omega^2$
resonantly matches one of the eigenfrequencies of the
stiffness matrix $K$.
In the absence of resonance, we introduce displacements
$\vec u = \vec x - \vec x^*$,
and find their equations of motion, 
\begin{align*}
  \ddot{\vec u} &= -K(\vec u + \vec x^* - \vec x_0) + 2\Omega \Gamma \dot{\vec u}
  +\Omega^2 \vec u + \Omega^2 \vec x^* \\
  &= -(K - \Omega^2\mathbb{1})\vec u + 2\Omega \Gamma \dot{\vec u}.
\end{align*}
Here, the stiffness matrix was shifted due to the centrifugal force,
and a new Coriolis term has appeared.
\par
In the following, we further assume
slow rotations and neglect the term proportional to $\Omega^2 \ll 1$.
This leads to the final equations of motion,
\begin{align}
  \ddot{\vec u} = -K \vec u + 2\Omega \Gamma \dot{\vec u}.
  \label{eq:dynamics-rot-final}
\end{align}
The Coriolis term proportional to $\dot{\vec u}$ is responsible for
breaking time-reversal symmetry in this classical system (the transformation
$t\mapsto -t$ maps $\dot{\mathbf{u}}$ to $-\dot{\mathbf{u}}$ but leaves all other terms invariant), 
analogous to the Lorentz force in a quantum electron gas~\cite{Hasan2010}.
Because the eigenmodes of Eq.~\eqref{eq:dynamics-rot-final} cannot be computed directly by inserting a harmonic ansatz,
we must resort to the equivalent first order system
\begin{align}
    \dot{\mathbf{y}} = D \mathbf{y}, \qquad D = \begin{pmatrix}
    0 & \mathbb{1} \\
    -K & 2 \Omega \Gamma
    \end{pmatrix},
    \label{e:eom-pre-se}
\end{align}
where $\mathbf{y} = (\mathbf{u}, \mathbf{w})^\top$, $\mathbf{w} = \dot{\mathbf{u}}$, and
$D$ is the dynamical matrix, the eigenmodes of which can be readily computed.

Equation~\eqref{e:eom-pre-se} can be brought into
a form manifestly equivalent to the Schr\"odinger equation
by introducing the change of variables~\cite{Susstrunk2016},
\begin{align*}
    \boldsymbol{\psi} = \begin{pmatrix}
    \sqrt{K} & 0 \\
    0 & i\mathbb{1}
    \end{pmatrix} \mathbf{y},
\end{align*}
where the matrix square root $\sqrt{K}$ is well-defined because
$K$ is positive-semidefinite. Under this change of variables, the dynamics becomes
\begin{align}
    i \dot{\boldsymbol{\psi}} = H \boldsymbol{\psi}, \qquad
    H =\begin{pmatrix}
    0 & \sqrt{K} \\
    \sqrt{K} & 2i \Omega \Gamma 
    \end{pmatrix},
    \label{eq:quantum-hamiltonian}
\end{align}
where the \lq Hamiltonian\rq{} $H$ is manifestly Hermitian.
This form makes explicit the connection between
classical mechanical and quantum systems, as now the
machinery of quantum mechanics is applicable to
Eq.~\eqref{eq:quantum-hamiltonian}.
\par
Below, we illustrate and analyze the generic consequences of time-reversal symmetry breaking via rotation for three distinct mechanical networks based on the inversely designed unit cells in Fig.~\ref{fig:1}. We will see that the corresponding MSNs 
undergo a topological phase transition when the rotation frequency exceeds a critical value, resulting in topologically protected gapless modes that are exponentially localized at the boundary of samples.

\section{Topological excitations in rotated networks}
The three mechanical networks from Fig.~\ref{fig:1} exemplify typical phenomena encountered with mechanical Chern networks.
For each of them, a topological phase transition occurs at some
finite $0 < |\Omega_c| < 0.1$, independent of the sign of $\Omega$.
This is due to the fact reversing the sign of the rotation
frequency $\Omega$ is equivalent to reversing time $t\mapsto -t$, and therefore mirrors the band structure, $\omega(\mathbf{k}, \Omega)=\omega(-\mathbf{k}, -\Omega)$.
In particular, this means that one can use the sign of $\Omega$ to  control the unidirectional propagation of excitations:  A wave packet will reverse direction when the sign of $\Omega$ is flipped.
In the topological phase $|\Omega| > \Omega_c$, all of the considered networks have a gap-Chern invariant $C = \pm 1$, which we calculated using the numerical procedure outlined in Ref.~\cite{Fukui2005}.

\begin{figure*}[t]
    \centering
    \includegraphics[width=\textwidth]{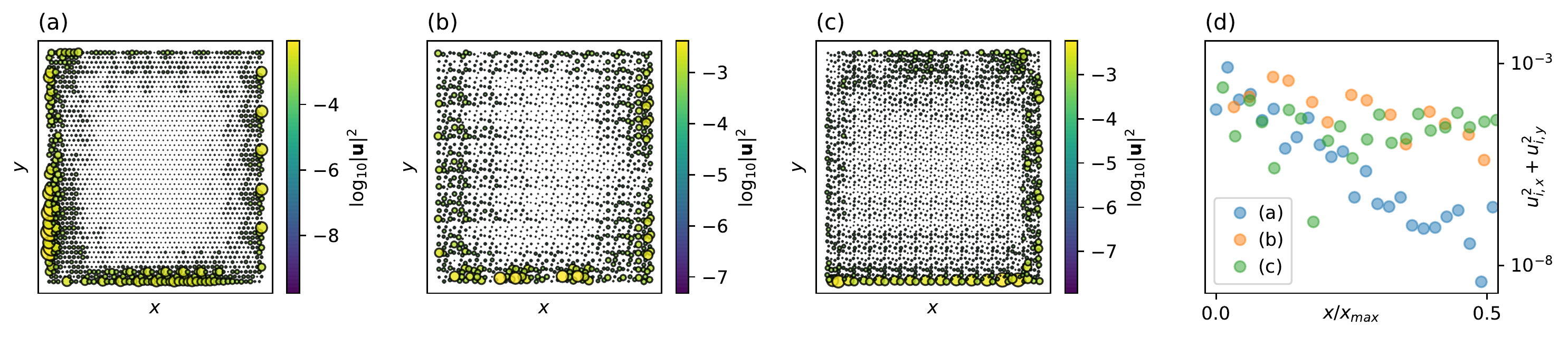}
    \caption{\textbf{Exponential localization of edge modes in
    rotating tuned mechanical Chern networks.} We construct finite realizations
    consisting of (a) $12\times 12$ and (b,c) $14\times 14$ unit cells
    of the three networks from Fig.~\ref{fig:1}, and plot
    a single eigenmode of the dynamical matrix $D$ from Eq.~\eqref{e:eom-pre-se} with a frequency inside the bulk gap.
    The values of $\Omega$ are $\Omega=0.1$ (a),
    $\Omega=0.15$ (b,c), in the 
    topological regime.
    The sizes and colors of the circles are proportional to the magnitude
    of the local node displacement
    $u_{i,x}^2 + u_{i,y}^2$, where $\mathbf{u} = (\mathbf{u}_x, \mathbf{u}_y)$
    is the eigenmode. Each network hosts topological modes entirely localized at the boundary.
    (d) We consider a small slice in $y$ direction of the networks from
    (a--c) and again plot the magnitude of the local node displacement.
    For all three networks, the magnitude decreases exponentially in the bulk,
    demonstrating localization of the modes to the boundary.
    }
    \label{fig:4}
\end{figure*}
\par
Generally, edge bands can be visualized by taking an
infinite periodic crystal in 2D and restricting to a ribbon-like  slice that is finite in one direction with open boundary conditions [Fig.~\ref{fig:2}(g)]. The resulting 1D crystal now possesses a one-dimensional band structure in which localized edge modes are directly visible. Mode localization can be measured by the participation ratio $\lambda = (\sum_n |u_n|^4)/(\sum_n |u_n|^2)^2$ of the eigenvector $\mathbf{u}(k)$. The ratio $\lambda$ is large if the mode is localized to few elements of the vector, and small if it is spread over many elements of the vector.
\par
For all three example networks from Fig.~\ref{fig:1}, the corresponding 1D crystals exhibit two bands of localized modes in the bulk gap in the topological phase $|\Omega| > \Omega_c$~[Fig.~\ref{fig:2}(b,d,f)]. The two bands host wave packets with opposite group velocity $v_g = d\omega/dk$, and are localized at opposite edges of the semi-infinite ribbon system.
They thus correspond to one single \emph{chiral} edge excitation.
The match between the bulk gap-Chern number $C=\pm1$ and the number of edge excitations (more precisely, the difference between clockwise and counter-clockwise edge modes) is a direct manifestation of the celebrated bulk-boundary correspondence~\cite{Hatsugai1993,Mong2011}.
\par
We further note that although the existence
of $|C|$ protected edge bands is guaranteed in the topological regime $|\Omega|>\Omega_c$, this
does not preclude unprotected edge states in the trivial phase $|\Omega|<\Omega_c$. To illustrate this fact explicitly, consider the example in Fig.~\ref{fig:2}(e,f). The band structure for  $\Omega=0$ in Fig.~\ref{fig:2}(e) is topologically trivial ($C=0$) but exhibits features two localized edge bands, which are converted into the topologically protected bands in Fig.~\ref{fig:2}(f) as one crosses the phase transition at finite $|\Omega|=\Omega_c > 0$.
\par
All three networks analyzed in Figs.~\ref{fig:1} and~\ref{fig:2} have in common that they support only a single chiral edge mode, the direction of which can be reversed by changing the sign of $\Omega$. Additional simulation scans suggest that this is typical of mechanical networks designed with the LRO scheme: Among all bandgap-designed networks that exhibited a topological transition, we never observed a case with $|C|>1$. This empirical finding is consistent with results from previous studies which reported that larger Chern numbers are typically associated with materials that possess long-range interactions or with systems that are periodically quenched or driven~\cite{Xiong2016}. Mechanical networks with long-range interactions could, in principle, be designed by introducing additional bonds that connect beyond the nearest neighbor unit cells. While certainly intriguing, such \lq non-local\rq{} networks are beyond the scope of the present study.

\par

The wave packets hosted by the topological edge bands of our short-range MSNs can be excited dynamically by forcing a semi-infinite or a finite network near the boundary at a frequency inside the bulk gap. As a specific example showcasing this generic effect, we consider the mechanical network from Fig.~\ref{fig:1}(c) and construct a finite realization consisting of $12\times 12$ unit cells. To demonstrate the robustness of the topological modes, we remove three unit cells from the left side boundary to introduce a boundary perturbation [Fig.~\ref{fig:3}(a)]. We then numerically simulate the forced dynamics
\begin{align}
\ddot{\mathbf{u}} + K \mathbf{u} - 2\Omega \Gamma \dot{\mathbf{u}} = \mathbf{f} 
\sin(\omega t) \, h(t),
\label{eq:dynamics-rot-final-forced}
\end{align}
where the forcing vector 
$\mathbf{f} = (1,  0, \dots, 0, 1,  0, \dots, 0)^\top$ is
zero except for the $x$ and $y$ components of one single node
near the bottom left corner. We pick $\Omega=0.15$ such that the network is
in the topological phase, and $\omega=0.71$ inside the bulk gap.
The window function \mbox{$h(t) = \sin(\pi t/150) \, \Theta(150 -t)$},
where $\Theta(t)$ is the Heaviside Theta function, slowly turns on the forcing
at $t=0$, and turns it off entirely at $t=150$.
The forcing injects energy into the network at the frequency $\omega$, which
preferentially excites edge modes and creates a wave packet
that travels unidirectionally along the edge of the network~[Fig.~\ref{fig:3}(b--e), Supplemental Movie 1].
In particular, due to the topological protection of the edge modes,
the precise shape of the boundary does not matter for the existence of
these wave packets. Back-scattering modes
are suppressed, and the wave packet is able to travel around the perturbation
in the boundary [Fig.~\ref{fig:3}(b,c)].
As anticipated at the beginning of this section, the chirality of these wave packets is controlled by the sign of the rotation rate~$\Omega$ (Supplemental Movie 2).
If the network is put in the topologically trivial regime,
no edge modes exist and the energy injected by forcing does not
create a chiral traveling wave packet~(Supplemental Movie 3).
\par
The dynamical behavior described above is
encoded in a set of eigenmodes $\mathbf{u}$ with
$D \mathbf{u} = i\omega\mathbf{u}$ that are exponentially localized to
the boundaries of the system, and where $\omega$ lies in the bulk gap.
For the three networks shown in Fig.~\ref{fig:1}, we again constructed finite realizations consisting of many unit cells in a square array, and computed the eigenmodes of the finite dynamical matrix $D$ from Eq.~\eqref{e:eom-pre-se}. For all three networks, we identified modes inside the bulk gap which were then found to be localized at the boundary [Fig.~\ref{fig:4}(a--c)].
To demonstrate exponential localization in each case, we analyzed
a slice of the eigenmodes in $y$ direction. Plotting
the logarithm of the average node displacement $u_{i,x}^2 + u_{i,y}^2$ as a function of the $x$ position of the node confirms an exponential decay of the node displacement with distance from the boundary~[Fig.~\ref{fig:4}(d)].

\section{Conclusions}

We have demonstrated the existence of topologically protected chiral edge modes in the gaps of \emph{inversely designed} mechanical networks, and have characterized their dynamical
properties. For the network realizations considered here, we found that band inversion near the gap was a robust predictor for a topological phase transition induced by sample rotation. The direction of rotation enables control over the chirality of the edge excitation, and topological protection of the edge excitations was confirmed in direct numerical simulations and through calculations of an appropriate Chern invariant.
\par
We hope that the present work can serve as a stepping stone towards the precise inverse programming of topological features into discrete disordered metamaterials. Instead of constructing gapped materials on the basis of known lattices by using certain features of the band structure (e.g., band inversion) as indicators of potential topological transitions, we envision that Linear Response Optimization~\cite{Ronellenfitsch2018} may eventually allow the direct tuning of such properties by implementing the desired topological characteristics into the optimization objectives.

\bibliography{references}

\end{document}